\PassOptionsToPackage{numbers,sort&compress}{natbib}
\documentclass{lhep}       

\makeatletter
\newif\ifalwaysfalse
\alwaysfalsefalse

\long\def\MaketitleBox{%
  \resetTitleCounters
  \let\@titlerunning\@title\ifalwaysfalse
  \else
  \vskip-18pt
  \hbox to \textwidth{Letters in High Energy Physics\hfill\@journal-\@vol,\ \@jyear}\par\vspace*{-8pt}\rlap{\rule{\textwidth}{1pt}}\vspace*{16pt}\fi
    \vbox{\centering{\titlefont\@title}}\vskip10pt
    \centering{\elsauthors}\par\vskip10pt
    {\adfont\elsaddress}\par
    \ifpublished\vskip-8pt 
    Published:~\@published\par\vskip14pt\else \vspace*{-0pt}\fi
    \ifvoid\absbox\else\unvbox\absbox\par\vskip10pt\fi
    \ifvoid\keybox\else\unvbox\keybox\par\vskip10pt\fi
    \date{}
  }
\makeatother

\makeatletter
\def\@evenhead{\ifpublished \else {Letters in High Energy Physics\hfil \@journal-\@vol,\ \@jyear}\par\llap{\rule[-4pt]{\textwidth}{1pt}}\fi}
\def\@oddhead{\ifpublished \else Letters in High Energy Physics\hfil \@journal-\@vol,\ \@jyear\par\llap{\rule[-4pt]{\textwidth}{1pt}}\fi}
\makeatother

\journal{LHEP}
\vol{301}
\jyear{2023}

\publishedtrue
\published{09 April 2023}

\usepackage[utf8]{inputenc}
\usepackage[english]{babel}
\usepackage[T1]{fontenc}
\usepackage{amsmath}

\usepackage{tikz}

\usepackage{graphicx}
\usepackage{dcolumn}
\usepackage{bm}
\usepackage[colorlinks=true,linkcolor=Blue,urlcolor=NavyBlue,citecolor=Fuchsia]{hyperref}

\def\be{\begin{equation}}
\def\ee{\end{equation}}

\newcommand{\beq}{\begin{equation}}
\newcommand{\eeq}{\end{equation}}
\newcommand{\bea}{\begin{eqnarray}}
\newcommand{\eea}{\end{eqnarray}}

\usepackage{multicol}

\begin{document}

\title{A Quantum Algorithm for Model-Independent Searches\\for New Physics}

\author{Konstantin T. Matchev,\auno{1} Prasanth Shyamsundar,\auno{1,2} and Jordan Smolinsky\auno{1}}
\address{$^1$Institute for Fundamental Theory, Physics Department, University of Florida, Gainesville, FL 32611, USA.}
\address{$^2$Fermilab Quantum Institute, Fermi National Accelerator Laboratory, Batavia, IL 60510, USA}

\begin{abstract}
  We propose a novel quantum technique to search for unmodeled anomalies in multidimensional binned collider data. We propose associating an Ising lattice spin site with each bin, with the Ising Hamiltonian suitably constructed from the observed data and a corresponding theoretical expectation. In order to capture {\em spatially correlated} anomalies in the data, we introduce spin-spin interactions between neighboring sites, as well as self-interactions. The ground state energy of the resulting Ising Hamiltonian can be used as a new test statistic, which can be computed either classically or via adiabatic quantum optimization. We demonstrate that our test statistic outperforms some of the most commonly used goodness-of-fit tests. The new approach greatly reduces the look-elsewhere effect by exploiting the typical differences between statistical noise and genuine new physics signals.
\end{abstract}

\maketitle

\begin{keyword}
goodness-of-fit tests\sep
anomaly detection\sep
quantum computing\sep
adiabatic quantum optimization\sep 
Ising spin model
\doi{10.31526/LHEP.2023.301}
\end{keyword}

\section{Introduction}
With the discovery of the Higgs boson \cite{Aad:2012tfa,Chatrchyan:2012xdj} at the CERN Large Hadron Collider (LHC), the Standard Model (SM) particle roster is complete 
and the search for new physics beyond the SM (BSM) is afoot. Given the many puzzles left unanswered by the SM (the flavor problem, the dark matter problem, and the $CP$ problem, to name a few), there is no shortage of ideas as to what that new physics may look like, yet we cannot be certain that the correct BSM theory has been written down and/or is being looked for by the current searches. This greatly motivates searching for BSM physics in a \emph{model-independent} way, as pioneered by the Tevatron and HERA experiments in the early 2000's \cite{Abbott:2000fb,Abbott:2000gx,Abbott:2001ke,Aktas:2004pz,Wessels:2007md,Aaltonen:2007dg,Aaltonen:2007ab,Aaltonen:2008vt,Aaron:2008aa,Abazov:2011ma} and pursued currently by the CMS and ATLAS LHC collaborations as well \cite{CMS:2008gya,Duchardt:2017mpl,Aaboud:2018ufy}.

The starting point in a typical BSM search is the prediction, obtained from Monte Carlo simulations, for the SM background in the relevant search regions in parameter space.\footnote{There are alternative approaches that try to avoid (to varying degrees) the reliance on a background prediction from Monte Carlo. These include traditional bump-hunting methods, edge detection techniques \cite{Debnath:2015wra,Debnath:2016mwb}, and recent machine-learning-based approaches \cite{Metodiev:2017vrx,Aguilar-Saavedra:2017rzt,Collins:2018epr,DeSimone:2018efk,Hajer:2018kqm,Heimel:2018mkt,Farina:2018fyg,Casa:2018avf,Cerri:2018anq,Collins:2019jip,Roy:2019jae,Dillon:2019cqt,Blance:2019ibf,Mullin:2019mmh,Nachman:2020lpy}. However, given the spectacular success of the SM in describing current data, its theoretical prediction of the background is well under control and should not be ignored. } The observed data, which can be in multiple bins or channels, is then compared to this expectation. The task of the experimenter is to test for consistency via some goodness-of-fit test \cite{Behnke:2013pga}. In this paper, we propose a novel, signal model-independent,  goodness-of-fit test, which takes into account not only the size of the observed deviations in the data but also their \emph{spatial correlations}. Real signals in the data are expected to exhibit strong spatial correlations, unlike statistical noise. For the purpose of quantifying the correlations, we introduce an Ising spin lattice with suitably defined nearest-neighbor spin-spin interactions (alternative approaches rely on neural networks \cite{DAgnolo:2018cun,DAgnolo:2019vbw} or wavelet transforms \cite{Contoyiannis:2019zmk,Lillard:2019exp,Beauchesne:2019tpx}).  Our proposed test statistic is the ground state energy ${\cal H}_{min}$ of the resulting Ising Hamiltonian ${\cal H}$. This method for anomaly detection greatly reduces the look-elsewhere effect and is very intuitive and easy to interpret. Finally, the proposed test statistic can be used not only for new physics searches but also for data quality monitoring and understanding the deficiencies of Monte Carlo event generation and detector simulation.

Finding the ground state of a general Ising model is a challenging computational problem, since analytical solutions exist only in very special cases \cite{Baxter}. For a relatively low number of bins, we can find ${\cal H}_{min}$ {\em exactly} by brute force, i.e., by examining all possible spin configurations. However, as the number of bins exceeds 30-40, this approach eventually becomes unfeasible, even with supercomputers, and one must resort to {\em approximate} classical methods like simulated annealing \cite{Kirkpatrick:1983zz}. Quantum computing algorithms also offer a promising avenue for solving such difficult combinatorial problems. The method of adiabatic quantum optimization (AQO) \cite{FGGS,FGGLLP,vDMV,Das:2008zzd} is particularly well suited for our problem, as it relies on the adiabatic theorem to find the ground state of a Hamiltonian of interest ${\cal H}$ as follows. One introduces a second Hamiltonian, ${\cal H}_0$, whose ground state is known in advance and easy to construct. At time $t=0$, the quantum system starts in the ground state of ${\cal H}_0$. Then, for a time $T$, the new Hamiltonian $H$ 
is smoothly interpolated as
\beq
H(t)=\left( 1-\frac{t}{T}\right) {\cal H}_0 + \frac{t}{T}\, {\cal H}.
\eeq
If $T$ is large enough, and ${\cal H}_0$ and ${\cal H}$ do not commute, the system remains close to the instantaneous ground state of $H(t)$. Then, measuring the ground state of $H$ at time $t=T$ returns an approximate solution to the original problem. This technique can be successfully applied to a large number of discrete optimization problems in applied mathematics, as long as one can find a Hamiltonian ${\cal H}$ whose ground state represents the desired solution (see \cite{Lucas} for a review). Examples of applications of quantum computing to problems in high energy physics include building a stronger Higgs classifier in the $\gamma\gamma$ channel~\cite{Mott:2017xdb}, unfolding distributions \cite{Cormier:2019kcq}, anomaly detection \cite{Alvi:2022fkk}, jet clustering \cite{Wei:2019rqy,Pires:2020urc,Pires:2021fka,deLejarza:2022bwc}, heavy flavor tagging \cite{Gianelle:2022unu}, track reconstruction \cite{Magano:2021jzd,Tuysuz:2021oai}, and many others \cite{Guan:2020bdl,Gray:2022fou}.

\section{The Ising model}
Our Hamiltonian is constructed as follows. Consider an arbitrary phase space of observable data partitioned into $N$ bins, labeled by $i=1,\ldots,N$. Each bin will be associated with a spin site $ s_i =\pm 1$ in our Ising lattice. Let $e_i$ and $o_i$ be the number of {\em expected} background events and {\em observed} events in the $i$-th bin, respectively. From those, we construct the corresponding normalized residuals\footnote{Note that the binning should be judiciously chosen so that the background expectations $e_i$ are large enough for all $i$, in order for the residuals to be well-behaved.}
\beq
 \Delta_i = \frac{o_i - e_i}{\sqrt{e_i}}, \qquad i=1,\ldots,N.
 \label{Deltadef}
\eeq
Note that in the background-only case, these normalized residuals are distributed according to the standard normal distribution ${\mathcal N}(0,1)$ which has no free parameters. Therefore, our results below will be insensitive to the size and shape of the background distribution.

Our goal in this letter is to build an Ising Hamiltonian ${\cal H}(\{\Delta_i\},\{s_i\})$ which depends on the set of measured deviations $\{\Delta_i\}$ and a set of spin configurations $\{s_i\}$ in such a way that the ground state energy ${\cal H}_{min}$ of the system is a measure of goodness-of-fit of the background hypothesis --- the lower the energy, the worse the fit. To this end, we define
\beq
\begin{split}
 &\mathcal{H}(\{\Delta_i\},\{s_i\}) \\
 &\qquad= -\sum_{i=1}^N \frac{|\Delta_i| \Delta_i}{2}\,\frac{s_i}{2} -\,\frac{\lambda}{2} \sum\limits_{i,j=1}^N w_{ij} \frac{(\Delta_i+\Delta_j)^2}{4}\,\frac{1 + s_i s_j}{2},
\end{split}
\label{eq:hamiltonian}
\eeq
\noindent
where $\lambda\ge 0$ is a free continuous parameter which controls the relative importance of the second term and $w_{ij}$ is a constant matrix which defines the range of spin-spin interactions. For simplicity, throughout this paper, we shall focus on nearest-neighbor\footnote{For the purposes of this paper, we shall define neighboring bins as follows: in two dimensions, they are two-dimensional tiles that share an edge; in three dimensions, they are three-dimensional solids that share a face, etc. } interactions only, where $w_{ij}=1$ if bins $i$ and $j$ are nearest neighbors, and $0$ otherwise. 

The ground state energy ${\cal H}_{min}$ is found by minimizing the Hamiltonian (\ref{eq:hamiltonian}) over the set $\mathcal{S}$ of all possible spin configuration sets $\{s_i\}$:
\beq
{\cal H}_{min}(\{\Delta_i\}) \equiv \min\limits_{\{s_i\}\in\mathcal{S}}\left\{\mathcal{H}(\{\Delta_i\},\{s_i\})\right\}.
\label{eq:teststatistic}
\eeq

The choice of Hamiltonian (\ref{eq:hamiltonian}) can be easily understood as follows. The first term is minimized when each spin $s_i$ is aligned with the corresponding deviation $\Delta_i$, i.e., when $s_i=\text{sgn}(\Delta_i)$. This means that in the $\lambda=0$ limit, where only the first term in (\ref{eq:hamiltonian}) survives, our test statistic (\ref{eq:teststatistic}) reduces to the familiar Pearson $\chi^2$ statistic (henceforth referred to simply as $\chi^2$):
\beq
\lim_{\lambda\to 0} {\cal H}_{min} = - \frac{1}{4} \sum\limits_{i=1}^N \Delta_i^2 = - \frac{\chi^2}{4}.
\label{limit}
\eeq

A well-known disadvantage of the $\chi^2$ test statistic is that it is insensitive to (a) the signs of the deviations $\Delta_i$ and (b) the relative proximity of the bins exhibiting the largest deviations (in absolute value). This is why it is desirable to complement the $\chi^2$ test with other, preferably independent, goodness-of-fit tests which would scrutinize the signs and the relative locations of the bins with the largest $|\Delta_i|$ \cite{GS}. One such test, applicable to one-dimensional binned distributions, is the Wald--Wolfowitz runs test~\cite{WW}, in which one inspects the series formed from the signs of the deviations
\beq
\text{sgn}(\Delta_1),\text{sgn}(\Delta_2),\text{sgn}(\Delta_3),\ldots,\text{sgn}(\Delta_N)
\label{signsofDeltas}
\eeq
and divides it into ``runs" --- successive nonempty strings of adjacent identical elements, i.e., where each string contains only plusses or only minuses. The number of such runs $r$ can be computed as
\beq
r = \frac{1}{2}\sum_{i=1}^{N-1} \left[1-\text{sgn}(\Delta_i)\text{sgn}(\Delta_{i+1})\right]
\label{rdef}
\eeq
and follows a Binomial distribution.
The runs test is less powerful than the $\chi^2$ test, since it does not use the magnitudes of the deviations. Nevertheless, it is still useful, since it is complementary to the $\chi^2$ test, and the two can be combined to form a more sensitive test; e.g., in Fisher's method, the product of the individual $p$-values is the new test statistic \cite{F1,F2}.

The second term in (\ref{eq:hamiltonian}) is designed to capture these effects in a more optimal way. We introduce interactions between neighboring spins, whose role is to try to align the spins among themselves --- the factor $(1 + s_i s_j)/2$ is designed to equal $1$ when the spins $s_i$ and $s_j$ are aligned and $0$ when they are anti-aligned. The effect is more pronounced if the deviations at the two neighboring sites are significant, i.e., both $\Delta_i$ and $\Delta_j$ are large in absolute value, and correlated, i.e., $\Delta_i$ and $\Delta_j$ have the same sign. It is precisely the effect of these latter correlations which we are trying to tap into in order to differentiate between random noise and meaningful physics signals. The effect of the second term in (\ref{eq:hamiltonian}) is controlled by the parameter $\lambda$, which for simplicity throughout this paper we shall fix to be equal to $1$.
\begin{figure}[t]
\centering
\includegraphics[width=0.99\columnwidth]{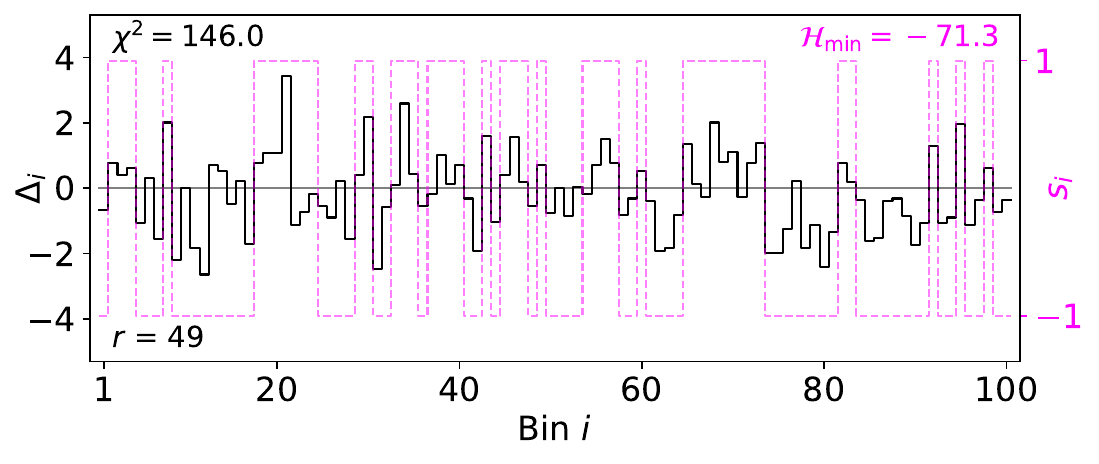}
\includegraphics[width=0.99\columnwidth]{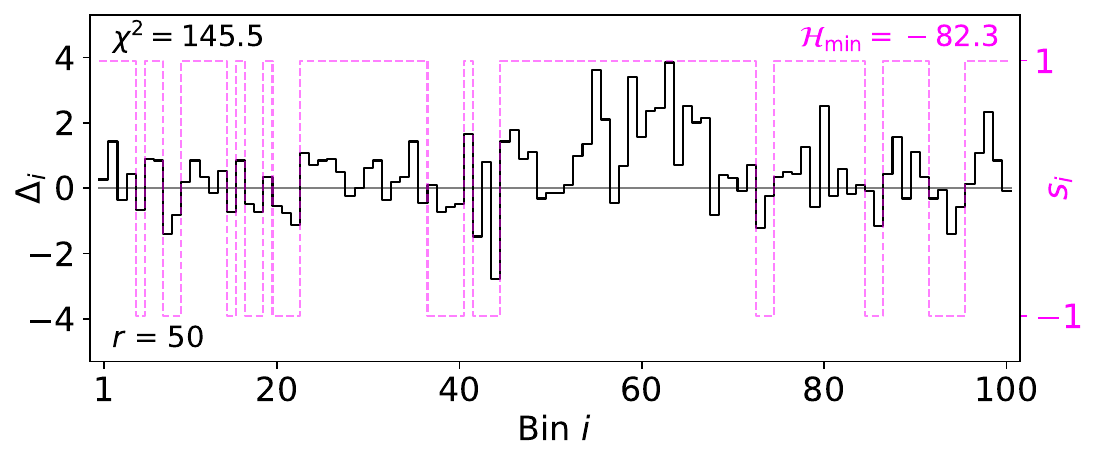}
\caption{Results from two representative pseudo-experiments with similar $\chi^2$ values: without signal (top) and with signal (bottom). The black solid histogram shows the deviations (\ref{Deltadef}), while the magenta dotted histogram depicts the corresponding spin configuration which minimizes the Hamiltonian (\ref{eq:hamiltonian}).}
\label{fig:qc1}
\end{figure} 

\section{Results with one-dimensional data}

In order to demonstrate the appropriateness of the Hamiltonian (\ref{eq:hamiltonian}), we first consider the following one-dimensional toy example illustrated in Figure~\ref{fig:qc1}. We take 100 equal-size bins which are populated with data sampled from a background distribution, which we take to be uniform, with an expected total number of 50000 events; and a signal distribution, which we take to be a normal distribution centered on the 60th bin with a standard deviation of 5 bin widths, and an expected total number of 500 signal events. In order to test the power of the ${\cal H}_{min}$ test, we generate 10,000 pseudo-experiments under the background hypothesis (top panel in Figure~\ref{fig:qc1}) and background plus signal hypothesis (bottom panel). For each pseudo-experiment, we first compute the resulting deviations (\ref{Deltadef}) shown with the black solid histogram and construct the Ising Hamiltonian (\ref{eq:hamiltonian}). Then, using the method of simulated annealing\footnote{For simplicity, we use a linear cooling schedule from $kT=10$ to $kT=0$ over 500000 steps. At each step, we calculate a Boltzmann factor associated with the transition to a new spin configuration differing by a single spin flip. These factors are then used to probabilistically decide on the new state of the system.} \cite{SimAnn}, we find the spin configuration $\{s_i\}$ (shown with the magenta dotted histogram) which minimizes the Hamiltonian and gives the ground state energy (\ref{eq:teststatistic}). Comparing the two types of histograms in Figure~\ref{fig:qc1}, we observe that the spins in the ground state indeed tend to align themselves in the regions where the deviations are strong and/or correlated, which is precisely what the Hamiltonian (\ref{eq:hamiltonian}) was designed to accomplish.

\begin{figure}[t]
\centering
\includegraphics[width=0.8\columnwidth]{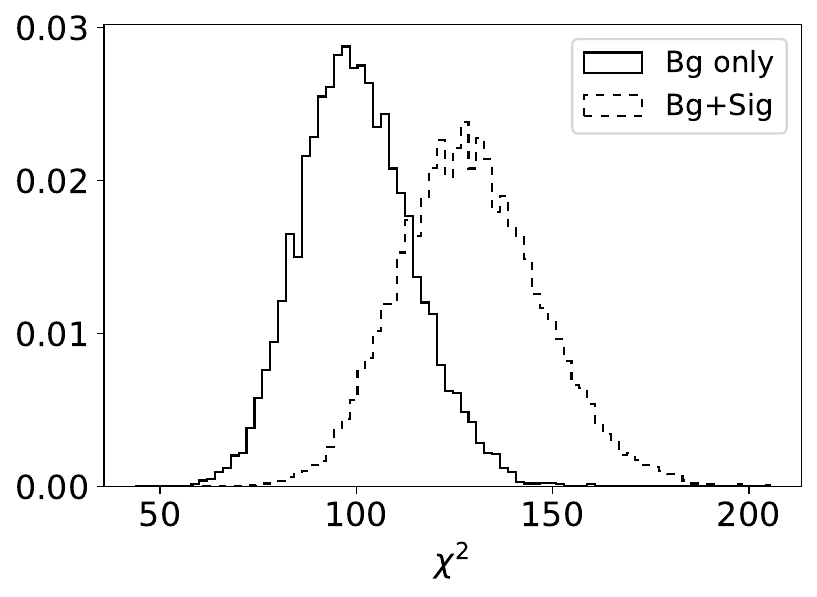}
\includegraphics[width=0.8\columnwidth]{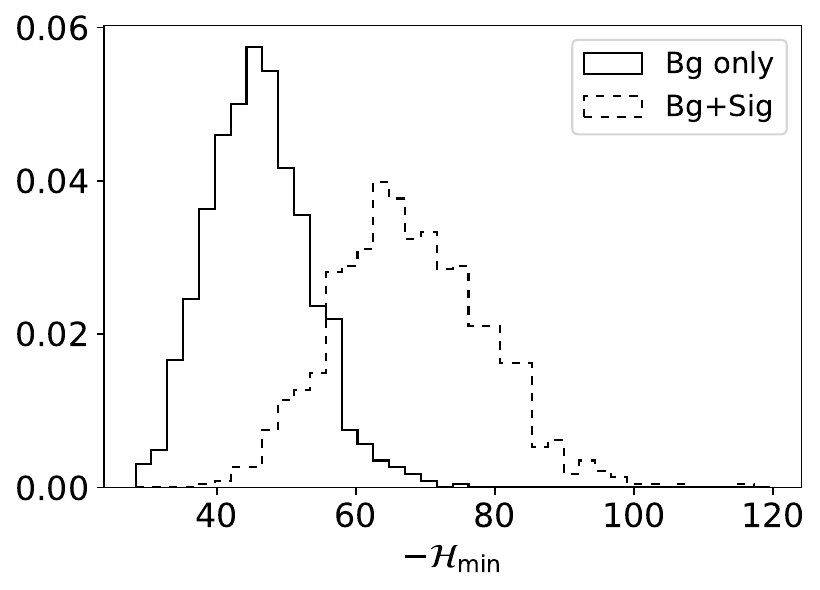}
\caption{Normalized distributions of the $\chi^2$ (top panel) and the $-{\mathcal H}_{min}$ (bottom panel) test statistics for the respective set of pseudo-experiments under the null (pure background) hypothesis (black solid lines) and the background plus signal hypothesis (black dashed lines).}
\label{fig:qc1hist}
\end{figure}

\begin{figure}[ht]
\centering
\includegraphics[width=0.75\columnwidth]{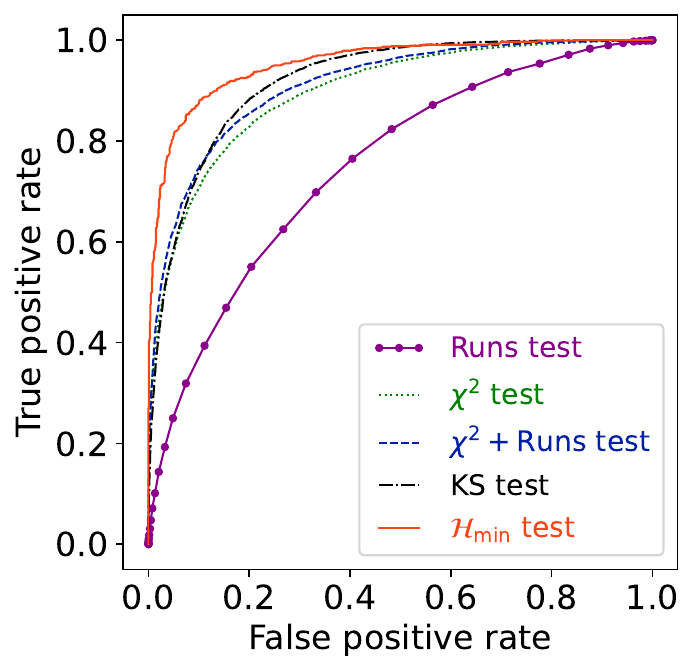}
\includegraphics[width=0.75\columnwidth]{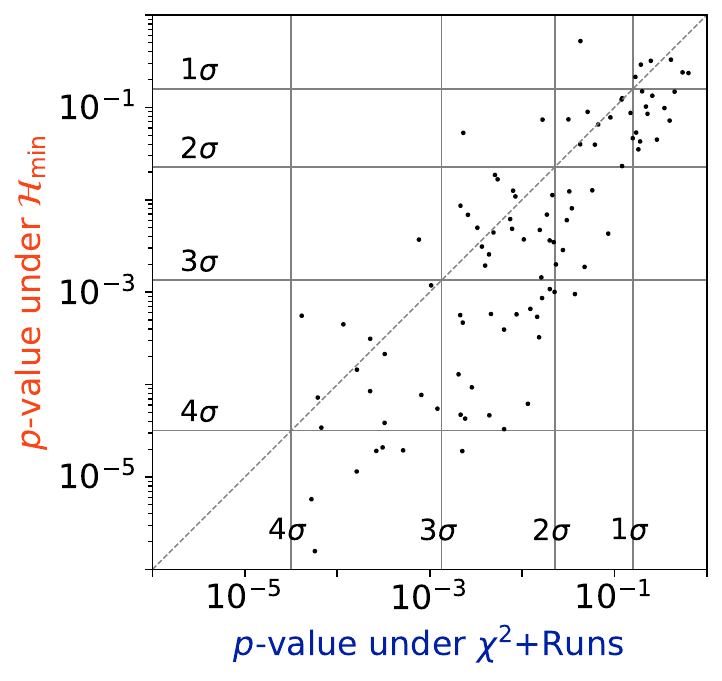}
\caption{Top: ROC curves for our ${\mathcal H}_{min}$ test statistic (orange solid line),
the Wald--Wolfowitz runs test (purple solid line with dot markers),
$\chi^2$-test (green dotted), combined $\chi^2$ plus runs test (blue dashed), and
the Kolmogorov-Smirnov test (black dot dashed).
Bottom: Scatter plot of the estimated $p$-values for 100 representative pseudo-experiments produced under the signal hypothesis, under the combined $\chi^2$ plus runs test ($x$-axis) and the ${\mathcal H}_{min}$ test statistic ($y$-axis).
The grid lines mark the $p$-values corresponding to a 1$\sigma$, 2$\sigma$, 3$\sigma$, and 4$\sigma$ effect.}
\label{fig:qc1roc}
\end{figure}

The two panels in Figure~\ref{fig:qc1} depict two pseudo-experiments whose $\chi^2$ values are rather similar,
146 and 145.5, respectively. Therefore, as far as the $\chi^2$ test statistic is concerned, these two sets of data appear very similar, even though the excess around 60 is rather evident to the trained eye. On the other hand, these two experiments produce rather different values for the ${\mathcal H}_{min}$ test statistic: $-71.3$ for the pure background case and $-82.3$ for the background plus signal case. 
This suggests that the ${\mathcal H}_{min}$ test statistic can perhaps better identify such signals in the data, but in order to verify this, we would need to look at the whole ensemble of pseudo-experiments. This more detailed comparison of the discriminating powers of the  $\chi^2$ and ${\mathcal H}_{min}$ test statistics is presented in Figures~\ref{fig:qc1hist} and \ref{fig:qc1roc}.
The top and bottom panels in Figure~\ref{fig:qc1hist} show the unit-normalized distributions of the $\chi^2$ and ${\mathcal H}_{min}$ test statistics, respectively, for large sets of pseudo-experiments under the background hypothesis (black solid lines) and the background plus signal hypothesis (black dashed lines). In order to accumulate enough statistics for the plots, in the top panel of Figure~\ref{fig:qc1hist}, we use data from 10,000 pseudo-experiments, while in the bottom panel we use only 1,000 pseudo-experiments due to the relative slowness of the ${\mathcal H}_{min}$ computation. In addition, in order to have the histograms similarly ordered from left to right on the two panels, in the bottom panel, we chose to plot $-{\mathcal H}_{min}$ instead of ${\mathcal H}_{min}$. We observe that in the case of the ${\mathcal H}_{min}$ test statistics (bottom panel) the distribution with the signal is further separated from the corresponding distribution for the null hypothesis, thus implying higher sensitivity.

In the top panel of Figure~\ref{fig:qc1roc}, we compare the sensitivity of our new test statistic to several standard goodness-of-fit tests
in terms of the corresponding receiver operating characteristic (ROC) curves \cite{ROC}. Results are shown for 
the new ${\mathcal H}_{min}$ test statistic (orange solid line),
the Wald--Wolfowitz runs test (purple solid line with dot markers),
$\chi^2$-test (green dotted), combined $\chi^2$ plus runs test (blue dashed), and
the Kolmogorov--Smirnov test \cite{GS} (black dot-dashed). It is clear that the new ${\mathcal H}_{min}$ test statistic outperforms all others, especially in the low false positive rate region which is relevant for discovery. The implications for discovery are further illustrated in the bottom panel of Figure~\ref{fig:qc1roc}, which shows a scatter plot of estimated $p$-values under the combined $\chi^2$ plus runs test ($x$-axis) and the ${\mathcal H}_{min}$ test statistic ($y$-axis), for 100 representative pseudo-experiments produced under the signal plus background hypothesis. We observe that for the large majority of the pseudo-experiments, namely, those below the $45^\circ$ dashed line, ${\mathcal H}_{min}$ gives a higher significance of discovery.
 
\begin{figure}[ht]
\centering
\includegraphics[width=0.45\columnwidth]{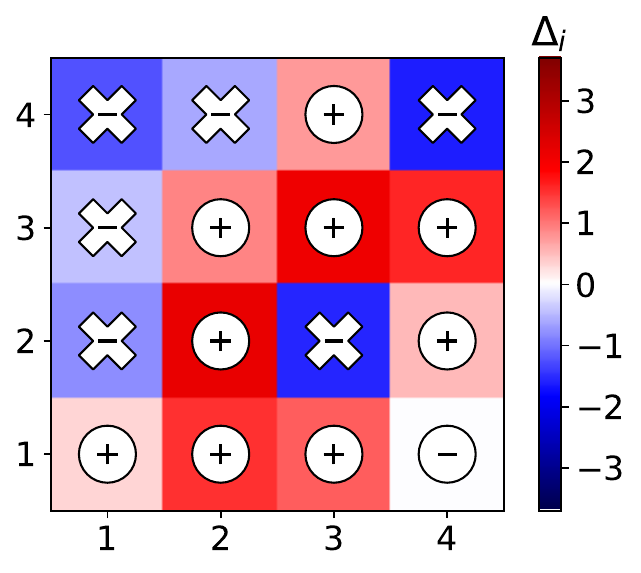}~~
\includegraphics[width=0.45\columnwidth]{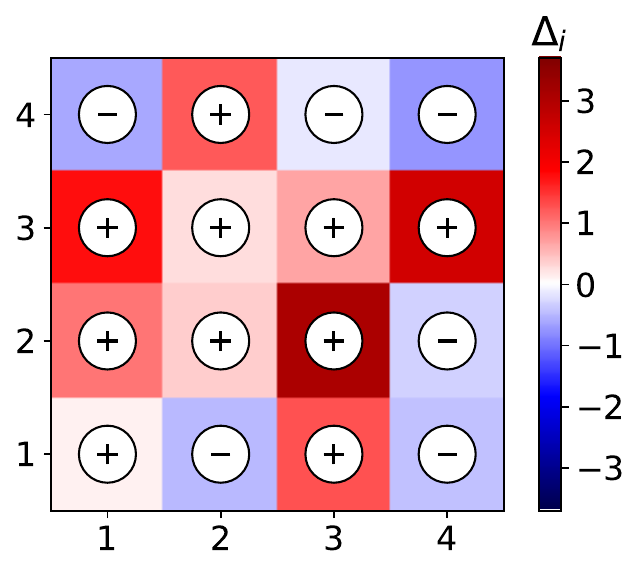}
\caption{Results from two representative pseudo-experiments with similar $\chi^2$ values for the $4\times 4$ two-dimensional exercise: without signal (left) and with signal (right), with the signal $2\times 2$ block located in the middle of the $4\times 4$ grid. Warm (cool) colors and plusses (minuses) indicate upward (downward) fluctuations (\ref{Deltadef}). Circles (crosses) indicate spin orientations $s_i=+1$ ($s_i=-1$) in the ${\mathcal H}_{min}$ spin configuration. }
\label{fig:qc2}
\end{figure} 
\begin{figure}[ht]
\centering
\includegraphics[width=0.45\columnwidth]{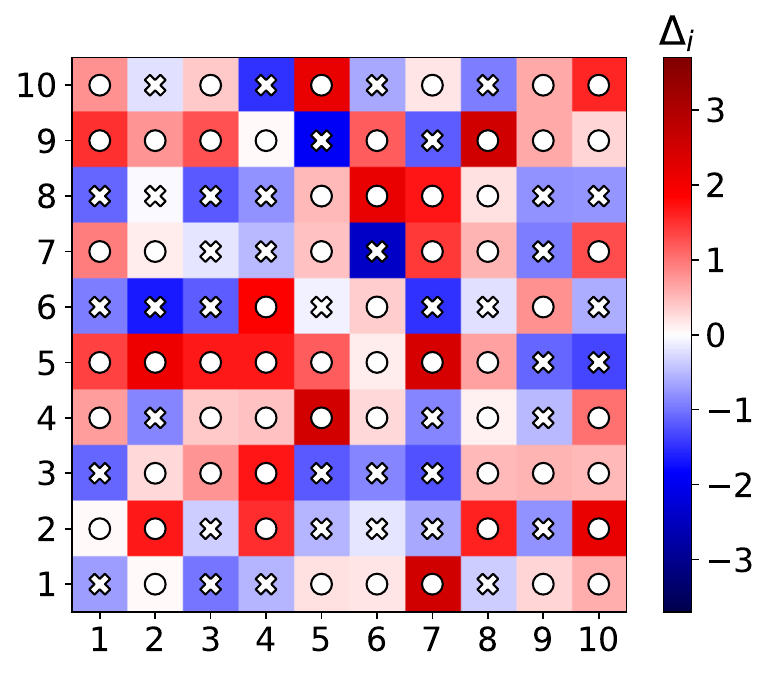}~~
\includegraphics[width=0.45\columnwidth]{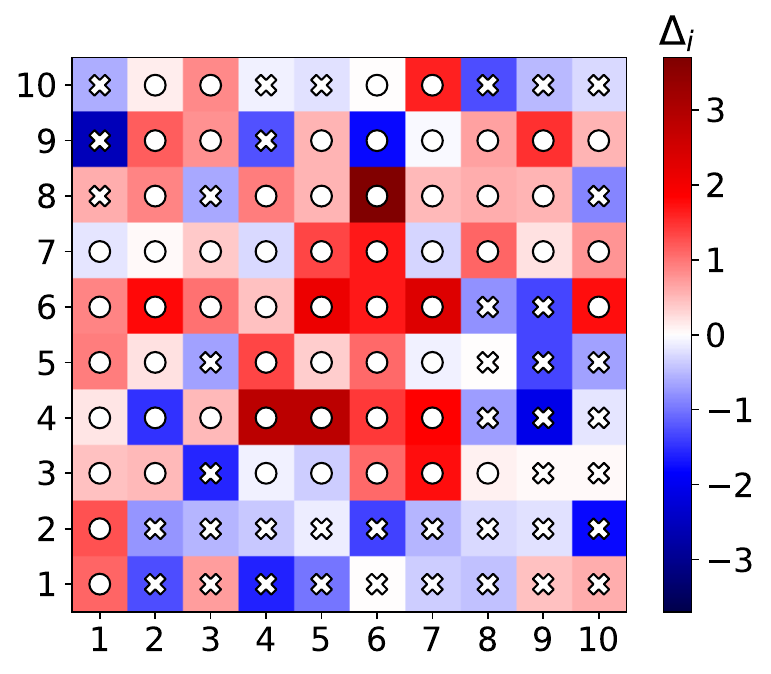}
\caption{The same as Figure~\ref{fig:qc2}, but for the $10\times 10$ exercise. The signal in the right panel is injected as an uncorrelated bivariate normal distribution with equal standard deviations of $1.5$ bin widths. Both pseudo-experiments have $\chi^2\approx129$.}
\label{fig:qc2-10}
\end{figure} 

\section{Results with two-dimensional data} 
\label{sec:results}

\begin{figure}[ht]
\centering
\includegraphics[height=6.5cm]{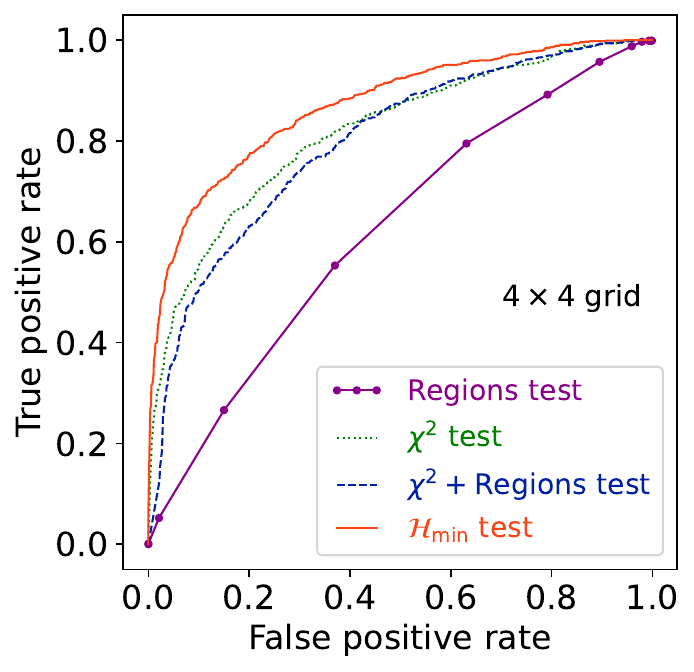}
\includegraphics[height=6.5cm]{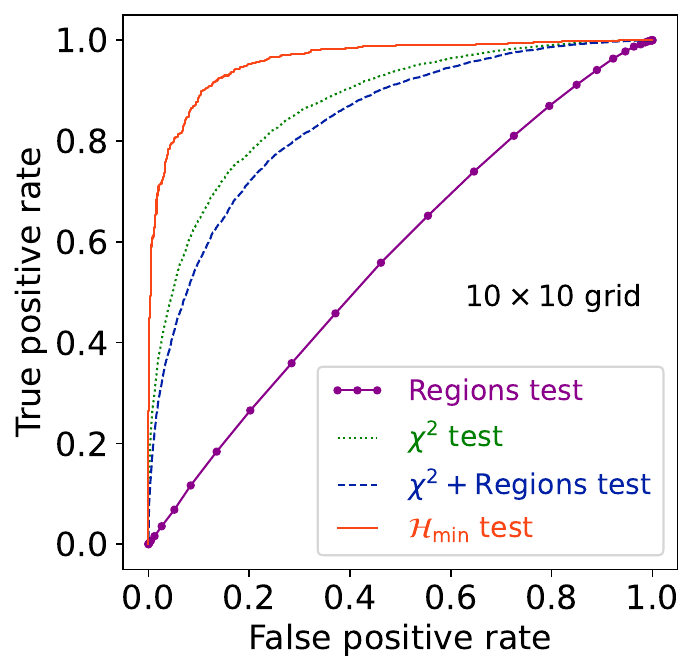}
\caption{ROC curves for the two-dimensional exercises with $4\times4$ data as in Figure~\ref{fig:qc2} (top) and $10\times10$ data as in Figure~\ref{fig:qc2-10} (bottom).}
\label{fig:qc2-ROC}
\end{figure}

Unlike existing tests sensitive to spatial correlations, our technique can be readily generalized to multidimensional data. This is illustrated in Figures~\ref{fig:qc2}\,--\,\ref{fig:qc2-ROC}, where for simplicity we limit ourselves to two dimensions and consider data arranged in an $n\times n$ grid of $N=n^2$ bins. 
In Figure~\ref{fig:qc2}, we take $n$ to be relatively low, $n=4$. This allows us to find the minimum energy ${\mathcal H}_{min}$ by brute force, i.e., by inspecting each of the $2^{N}=2^{16}$ spin configurations and comparing the corresponding energies. Then, in Figure~\ref{fig:qc2-10}, we consider a larger grid with $n=10$, for which the brute force method is unfeasible, and in order to find ${\mathcal H}_{min}$, we resort back to the method of simulated annealing used in the earlier one-dimensional example. In the $4\times4$ case of Figure~\ref{fig:qc2}, the $\Delta_i$ values for the background are sampled directly from the standard normal distribution ${\mathcal N}(0,1)$, and the signal is then modeled as a constant $1.5\sigma$ additional contribution to each bin in a $2\times2$ block of the $4\times4$ grid; i.e., the $\Delta_i$ values for the $2\times2$ block are sampled from ${\mathcal N}(0,1)+1.5$. The color code used in Figure~\ref{fig:qc2} reflects the resulting $\Delta_i$ values for the two pseudo-experiments --- upward (downward) fluctuations $\Delta_i$ are represented with warm (cool) colors and marked with plus (minus) signs. In the $10\times10$ case of Figure~\ref{fig:qc2-10}, the data values $o_i$ for the background are sampled from a Poisson distribution with $e_i=500$ in each bin. An uncorrelated bivariate normal signal (with equal standard deviations of $1.5$ bin widths) of 600 total expected events is then injected at the location of the $(6,6)$ bin. The resulting deviations $\Delta_i$ are then computed and shown for two representative pseudo-experiments in Figure~\ref{fig:qc2-10}, using the same color code as in Figure~\ref{fig:qc2}. Note that the index $i$ in eqs.~(\ref{Deltadef}-\ref{limit}) is now two-dimensional and identifies the horizontal and vertical location of the respective bin; two bins are considered nearest neighbors only if they share an edge.
\begin{figure}[ht]
\centering
\includegraphics[width=0.8\columnwidth]{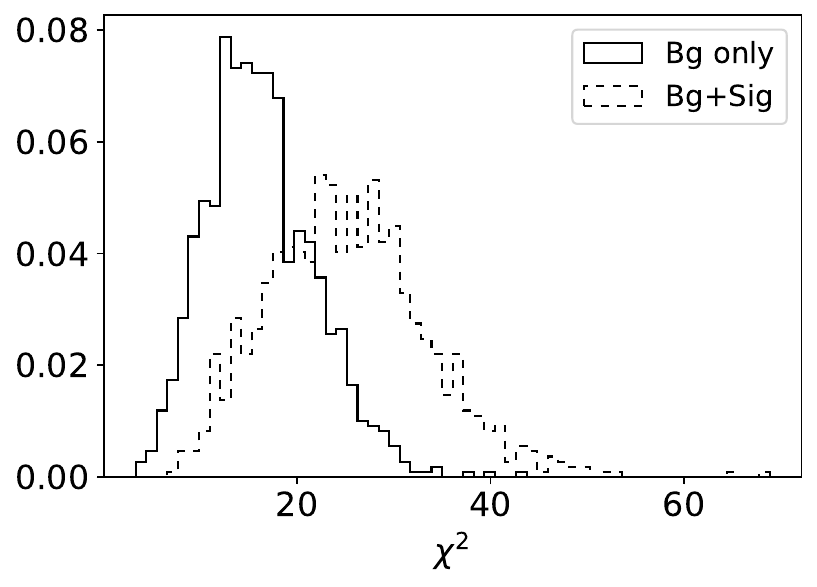}
\includegraphics[width=0.8\columnwidth]{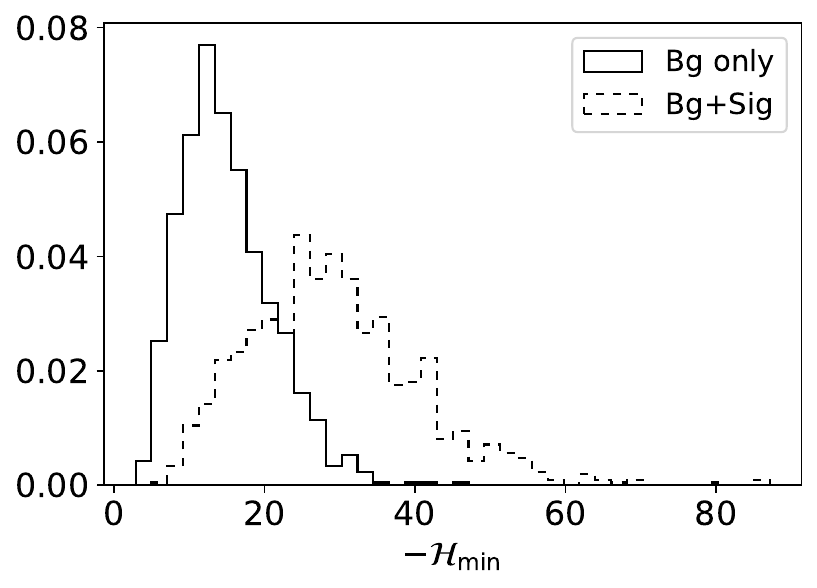}
\caption{The same as Figure~\ref{fig:qc1hist}, but for the $4\times 4$ two-dimensional exercise illustrated in Figure~\ref{fig:qc2}, using 1,000 pseudo-experiments.}
\label{fig:qc2hist}
\end{figure}
\begin{figure}[ht]
\centering
\includegraphics[width=0.8\columnwidth]{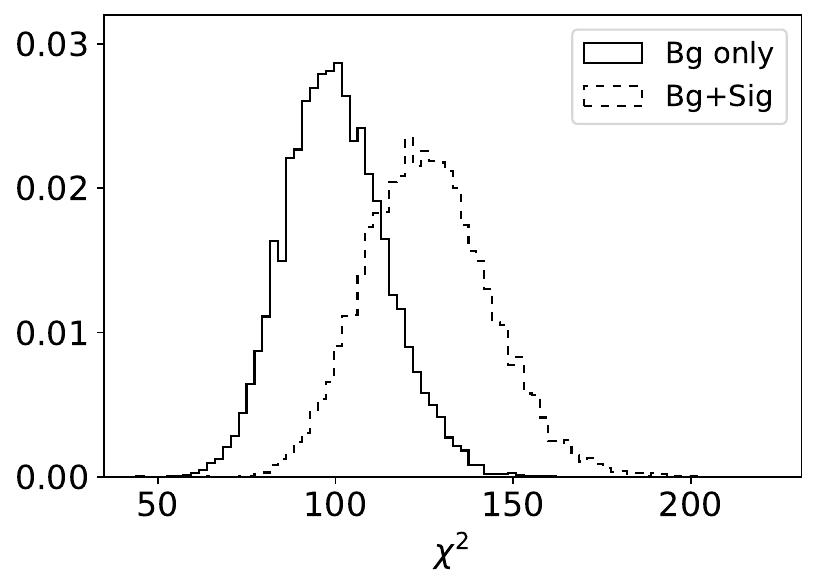}
\includegraphics[width=0.8\columnwidth]{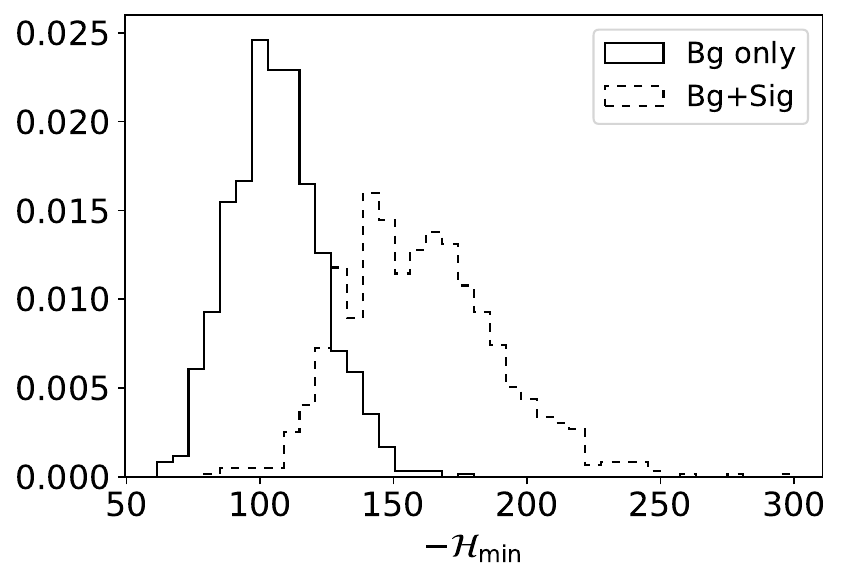}
\caption{The same as Figure~\ref{fig:qc2hist}, but for the $10\times 10$ two-dimensional exercise illustrated in Figure~\ref{fig:qc2-10}, using 10,000 pseudo-experiments for $\chi^2$ and 1,000 for $\mathcal{H}_\mathrm{min}$.}
\label{fig:qc2-10hist}
\end{figure} 

In complete analogy to Figure~\ref{fig:qc1hist}, in Figures~\ref{fig:qc2hist} and \ref{fig:qc2-10hist}, we show the respective distributions in our pseudo-experiments of the $\chi^2$ and $-{\mathcal H}_{min}$ test statistics for each of the two-dimensional examples considered in this section. We notice that, just like in the case of the one-dimensional exercise depicted in Figure~\ref{fig:qc1hist}, the ${\mathcal H}_{min}$ test statistic offers better separation of signal and background. We can quantify this separation in terms of the overlap area between the two distributions --- in Figure~\ref{fig:qc2hist}, it is 51.3\% in the case of the $\chi^2$ test statistic, but only 41.4\% in the case of ${\mathcal H}_{min}$. The improvement is even better in the case of the $10\times 10$ grid --- in Figure~\ref{fig:qc2-10hist}, the overlap area is 40.7\% for the $\chi^2$ test statistic and is reduced to only 20.1\% in the case of ${\mathcal H}_{min}$.

Motivated by the usefulness of the runs test in the one-dimensional example of Figures~\ref{fig:qc1} and \ref{fig:qc1roc}, we can attempt to generalize it to the two-dimensional data of Figures~\ref{fig:qc2} and \ref{fig:qc2-10}. For example, we can count the number of connected ``regions'' of only positive or only negative deviations $\Delta_i$, defined so that the nearest neighbors with the same $\text{sgn}(\Delta_i)$ necessarily belong to the same region. In that case, each of the two pseudo-experiments in Figure~\ref{fig:qc2} leads to 5 regions, as can be easily seen by inspecting the plusses and minuses shown in the bins. However, the so-defined ``regions'' test statistic is not very powerful, as can be seen from the respective ROC curves in Figure~\ref{fig:qc2-ROC} --- in fact, combining the ``regions'' test with the $\chi^2$ statistic generally makes things worse than using $\chi^2$ alone.

This is where the new test statistic ${\mathcal H}_{min}$ comes to the rescue. Figures~\ref{fig:qc2} and \ref{fig:qc2-10} depict the spin configurations in the respective ground states: circles indicate spin orientation $s_i=+1$ while crosses correspond to $s_i=-1$. The corresponding ROC curves in Figure~\ref{fig:qc2-ROC} (orange solid lines) demonstrate the superior performance of the ${\mathcal H}_{min}$ test statistic for these two-dimensional examples as well.

\begin{figure*}
 \centering
 \includegraphics[height=.19\textwidth]{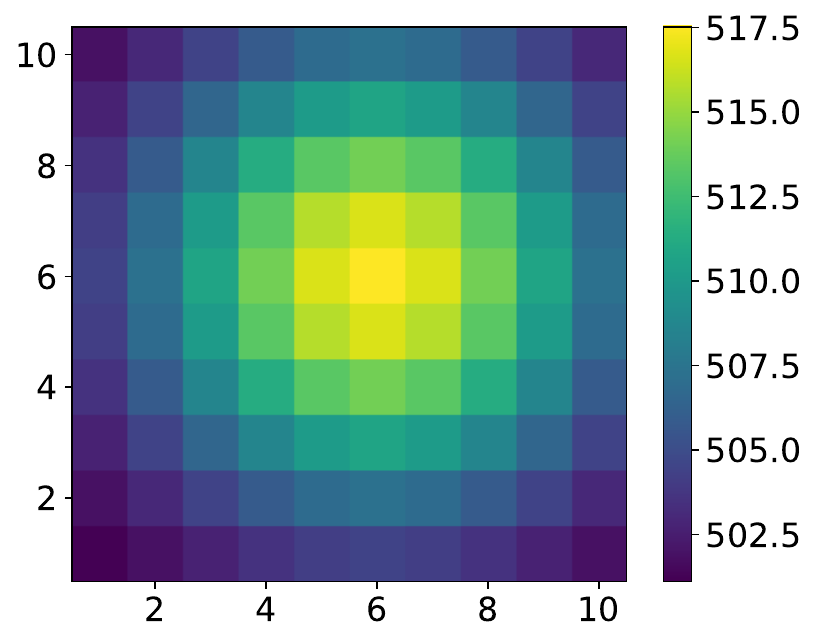}
 \includegraphics[height=.19\textwidth]{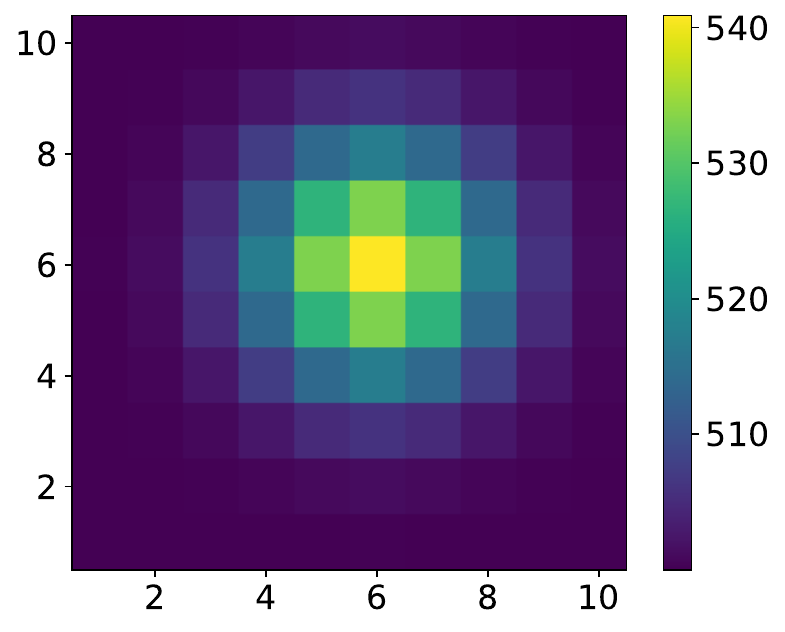}
 \includegraphics[height=.19\textwidth]{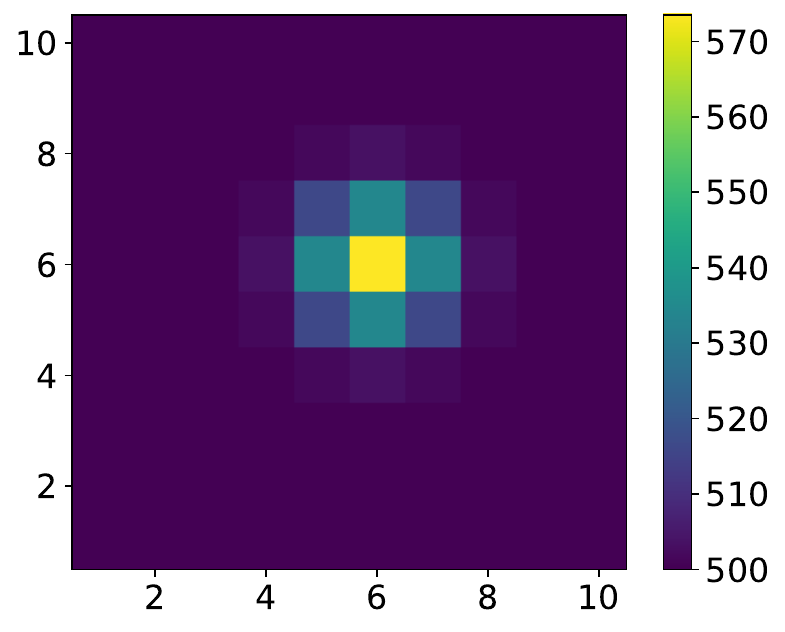}
 \includegraphics[height=.19\textwidth]{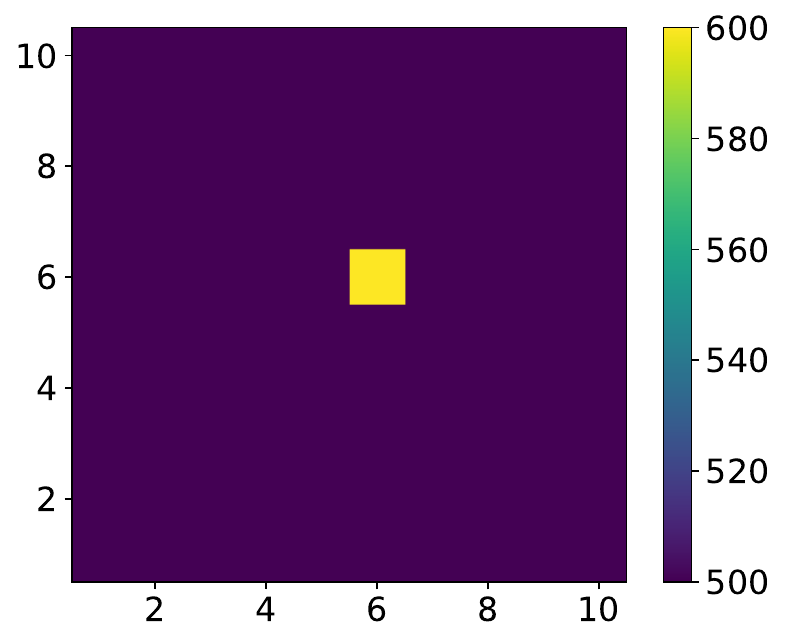}\\
  \includegraphics[height=.21\textwidth]{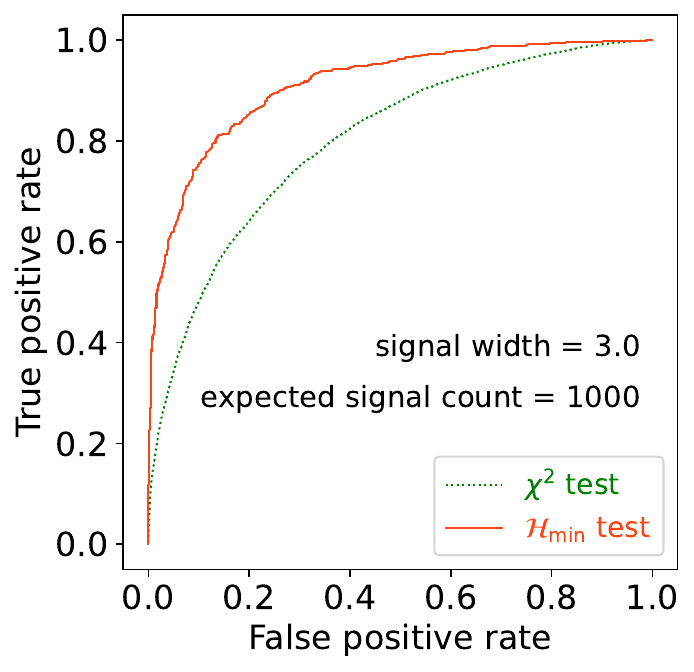}~~~~~~
 \includegraphics[height=.21\textwidth]{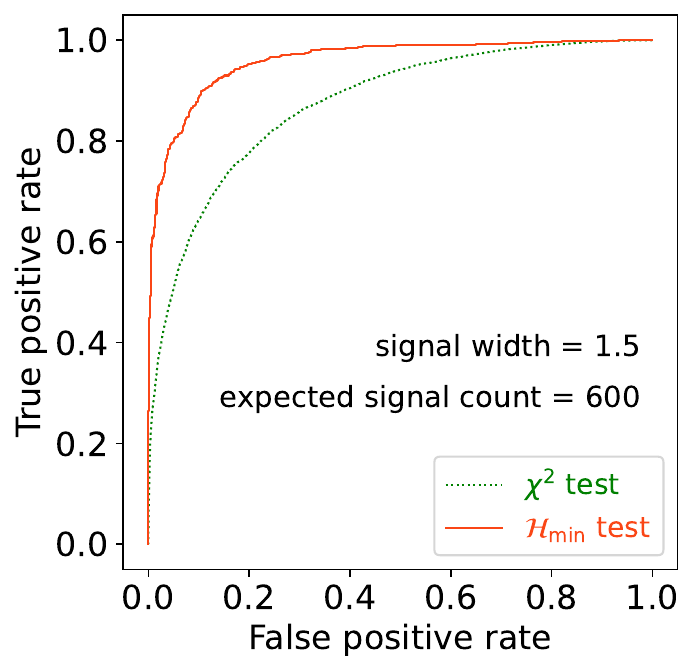}~~~~
 \includegraphics[height=.21\textwidth]{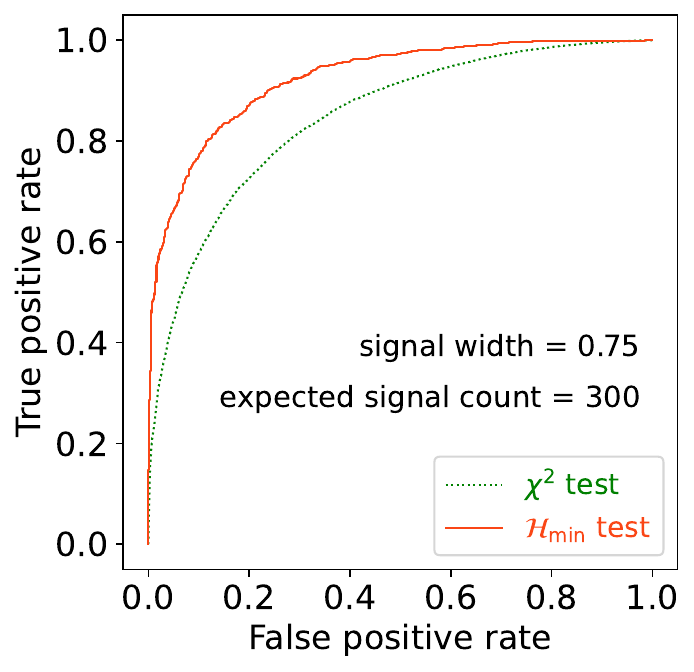}~~~~
 \includegraphics[height=.21\textwidth]{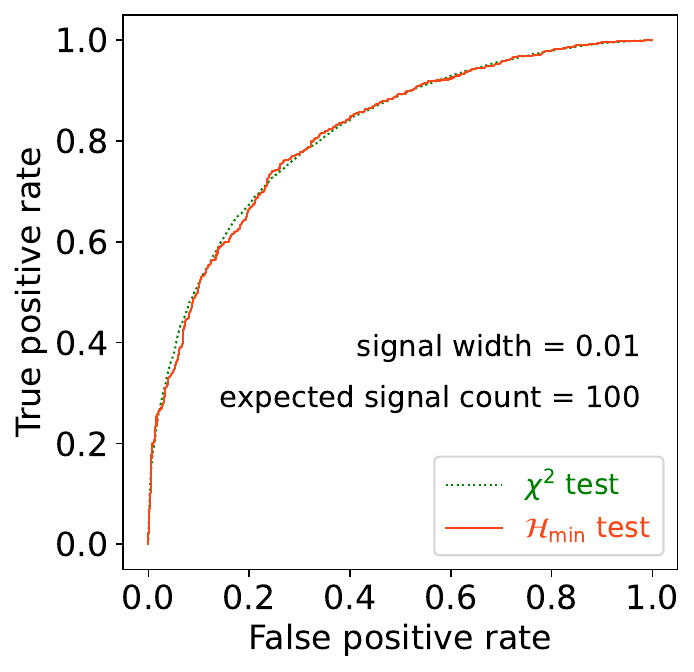}~~~~~~~~~
 \caption{Top row: heatmaps of the expected bin counts for the combined signal plus background distributions in the four examples considered at the end of Section~\ref{sec:results}. The background is flat with an expectation of 500 events per bin. The signals are bivariate normal distributions centered on the (6,6)-bin with standard deviations for both variates being (from left to right) 3.0, 1.5, 0.75, and 0.01 bin widths, respectively. Bottom row: ROC curves for these four exercises for the $\chi^2$-test (green dotted lines) and the $\mathcal{H}_\mathrm{min}$ test (red solid lines). The total expected signal counts are (from left to right) 1000, 600, 300, and 100, respectively.}
 \label{fig:8panels}
\end{figure*}

We note that our method is agnostic about the expected shape of the signal. However, its performance will, of course, depend on the properties of the signal distribution, e.g., its shape, location, width, etc.\footnote{No hypothesis test can be universally better for all signal models. For the same chosen false positive rate threshold, different hypothesis tests will be better for different signal models.} To illustrate this, we perform a series of exercises in which we vary the width of our previously injected signal as shown in the top four panels in Figure~\ref{fig:8panels}. Each panel is a heatmap of the expected bin count for the combined signal plus background distribution. As before, the background is flat with an expectation of 500 events per bin. The signals are bivariate normal distributions centered on the (6,6) bin with standard deviations for both variates being (from left to right) 3.0, 1.5, 0.75, and 0.01 bin widths, respectively. The total expected signal counts are (from left to right) 1000, 600, 300, and 100, respectively. Note that in the fourth column of Figure~\ref{fig:8panels}, all the signal events are roughly confined within one bin.

The corresponding ROC curves for these four exercises (for the $\chi^2$-test and the $\mathcal{H}_\mathrm{min}$ test) are shown in the bottom panels of Figure~\ref{fig:8panels}. Note that when the signal is localized (fourth column), the performance of the $\chi^2$-test roughly matches that of our $\mathcal{H}_\mathrm{min}$. This is understandable, since when the signal is confined to a single bin, there is no signal shape to speak of, and any correlated excesses observed in the data are due to random fluctuations, which nullifies any advantage of our method. On the other hand, as long as the signal extends over a few neighboring bins (left three columns), the correlations are captured by our method and this results in improved sensitivity.

\section{Conclusions and outlook}
In this paper, we proposed a new goodness-of-fit test statistic, ${\mathcal H}_{min}$, to identify deviations from an expectation, {\em without assuming any alternative hypothesis} to account for the deviations. Our test statistic exploits in a novel, model-independent way, the spatial correlations in the observed fluctuations of binned data relative to a theoretical prediction.  Our method for anomaly detection greatly reduces the look-elsewhere effect by exploiting the typical differences between the properties of statistical noise and real new physics effects. With several toy examples, we demonstrated that the ${\mathcal H}_{min}$ test performs better than some commonly used goodness-of-fit tests. Once a signal is detected, the spin configuration in the ground state can be inspected to identify atypically large domains of aligned spins which can then be used to interpret the origin of the anomaly detected by our statistic.

When an experiment calls for an analysis with a large number of bins (i.e., lattice spin sites) $N$, the {\em exact} computation of ${\mathcal H}_{min}$ becomes intractable and must be handled via a suitable approximate method. A promising approach to tackle large $N$ cases is offered by current and future AQO implementations on quantum computers \cite{Chimera,Albash:2017hpt,DSC}. In the meantime, an acceptable\footnote{Existing benchmark studies show that classical methods scale roughly on par with {\em current} adiabatic quantum optimizers \cite{Boixo,Ronnow,Parekh}.} alternative is to apply approximate stochastic optimization methods like simulated annealing, which was used in our analysis.

In this paper, we assumed that the theoretical expectation $\{e_i\}$ is known exactly. However, in realistic situations, e.g., in the presence of systematic uncertainties, it may depend on various nuisance parameters $\vec{\theta}$, in which case the test statistic can be modified as 
\beq
\max_{\vec{\theta}} \left\{ {\mathcal H}_{min}\left(\vec\theta\right) \right\}.
\eeq
 We are also in the process of exploring a larger class of Hamiltonians and their relevance to various combinatorial optimization problems, both inside and outside particle physics. We believe this work is only scratching the surface of a very interesting new direction of interdisciplinary research bridging condensed matter physics (Ising models), quantum information science, computational geometry, statistics, and high energy physics. 

\section*{Conflicts of interest} The authors declare that there are no conflicts of interest regarding the publication of this paper.

\section*{Acknowledgments} The authors would like to thank D.~Acosta, I.~Furic, S.~Hoffman, P.~Ramond, J.~Taylor, W.~Xue, and especially S.~Mrenna for useful discussions. The work of KM and PS is supported in part by the US Department of Energy under Grant No.~DE-SC0010296. The work of PS is supported by the University of Florida CLAS Dissertation Fellowship funded by the Charles Vincent and Heidi Cole McLaughlin Endowment. PS is grateful to the LHC Physics Center at Fermilab for hospitality and financial support as part of the Guests and Visitors Program in the summer of 2019.

\end{document}